# Quantum Processes
Jason Smith

*Abstract*

The exact scattering solutions of the Klein-Gordon equation in cylindrically symmetric field are constructed as eigenfunctions of a complete set of commuting operators. The matrix elements and the corresponding differential scattering cross-section are calculated. Properties of the pair production at various limits are discussed.

## *I. Introduction*

Since the pioneering work of Aharonov and Bohm [1] about half a century ago, the systems in which charged particles interact with the vector potential of an infinitely long, thin magnetic string (AB potential) are still receiving considerable interest in the literature. In such systems, the non-local interaction of the charged particle with the magnetic field of the string leads, quantum mechanically, to observable physical effects despite the absence of Lorentz forces on the particle. The works [2, 3] provide an excellent review of the subject and its application in various areas.

Recently, works addressing issues other than the elastic scattering of charged particles off the AB potential, had appeared. Serebryany *et al* [4] reported the differential cross section for Bremsstrahlung of non-
*et al* [5] considered the synchrotron radiation by a relativistic scalar particle in the AB potential. Bremsstrahlung by spin-1/2 particles was considered in [6], and the cross section for the electron-positron pair production by a single photon in the AB potential was calculated in [7].

The present work was mainly motivated by Skarzhinski *et al* [6-8]. In this work, we consider the production of a scalar particle-antiparticle pair in an AB potential by a single, linearly polarized photon. The exact differential cross section for the process is calculated and the limits of low and high-energy photons are discussed. Selection rules on the angular momentum of the resulting pair, that were reported for fermions [6,7], are also revealed in the scalar case. The framework of our calculations is the covariant perturbation theory. However, we expand the scalar field operators using, as basis, the exact scattering particle-antiparticle solutions of the Klein-Gordon (KG) equation in the AB potential, rather than using the free solutions of the free KG equation as basis as usually done (for the Coulomb potential, for instance) [9-11]. Such an approach is discussed in [12] and is used in [6,7]. This, in turn, makes the lowest non-vanishing contribution to the process of the first order rather than second order. Calculations of the differential cross section of Bremsstrahlung by a single photon, in the AB potential, using the free particle solutions as expansion basis were also carried out and will be published elsewhere.

In Section II, we construct the particle and antiparticle solutions of the KG equation with coupling to the AB potential. The calculation of the differential cross-section for the scalar particle-antiparticle pair by a single, linearly polarized photon is carried out in Section III. Section IV discusses the limits for low and high-energy photons. Finally, we sum up and state our conclusions in Section V.

## *II. The Exact Scattering Solutions to the Klein-Gordon Equation in the AB Potential*

In the presence of an external AB field, specified by the vector potential, one should make use of the minimal coupling in which $\hat{P}_m \to \hat{P}_m - \frac{e}{c} A_m$ (with e = -|e|). In this case, the interacting Klein-Gordon equation becomes:

$$(\hat{P}^m - \frac{e}{c} A^m)(\hat{P}_m - \frac{e}{c} A_m)\Psi(\vec{r},t) = M^2 c^2 \Psi(\vec{r},t) \qquad (1)$$

In this paper we will consider only an idealized case (pure AB case) of an infinitely thin, infinitely long straight magnetic tube. In the absence of Coulomb potential and in cylindrical coordinates (ρ, φ, z), it can be readily shown that the external vector potential has a nonzero angular component

$$eA_\varphi = \frac{e\Phi_{en}}{2\pi r} = \frac{-\Phi_{en}\hbar c}{\Phi_0 \; r} = \frac{-\hbar c}{r} f \qquad (2)$$

where $\Phi_{en}$ $\Phi_0$ is the enclosed magnetic flux,

$\Phi_0 = 2\pi\hbar c/e$ is the magnetic flux quantum,

$f = [f] + \delta$ with $[f]$ represents the integer part of $f$, and $\delta$ is fractional quantity that produces all physical effects.

Next, substituting for the vector potential into Eq.(1), then in the Coulomb Gauge, it reads

$$\left[\frac{1}{c^2}\frac{\partial^2}{\partial t^2} - \frac{1}{r}\frac{\partial}{\partial r}r\frac{\partial}{\partial r} - \frac{1}{r^2}\frac{\partial^2}{\partial \varphi^2} - \frac{\partial^2}{\partial z^2} + \frac{M^2 c^4}{\hbar^2} + \frac{f^2}{r^2} - \frac{2if}{r^2}\frac{\partial}{\partial \varphi}\right]\Psi(\vec{r},t) = 0 \qquad (3)$$

However, $\Psi(\vec{r},t)$ is an eigenfunction of the following operators: the third component of linear momentum $\hat{P}_3$, the third component of the angular momentum $\hat{L}_3$, and the Hamiltonian $\hat{H}$:

$$\hat{P}_3 \Psi(\vec{r},t) = k_3 \hbar \Psi(\vec{r},t) \, , \; \hat{L}_3 \Psi(\vec{r},t) = m\hbar \Psi(\vec{r},t) \, , \; \hat{H}\Psi(\vec{r},t) = E_n \Psi(\vec{r},t) \qquad (4)$$

where $k_3$, $m$ and $E_n = c\varepsilon_n$ are the eigenvalues of $\hat{P}_3$, $\hat{L}_3$ and $\hat{H}$ operators, respectively. In other words, $\{\hat{P}_3, \hat{L}_3, \hat{H}\}$ constitutes a complete set of commuting operators that are integrals of the motion. In this *ansatz*, the partial mode solution for $E_n > 0$, with normalization constant $N$, reads

$$\psi_{(+)}(r,\varphi,z;t) = N e^{-i(c\varepsilon_n t/\hbar - k_3 z - m\varphi)} R_m(r) \qquad (5)$$

where $R_m(r)$ obey the following radial KG equation:

$$\left[\frac{d^2}{dr^2} + \frac{1}{r}\frac{d}{dr} + k_\perp^2 - \frac{(f+m)^2}{r^2}\right] R_m = 0. \qquad (6)$$

Eq.(6) is the usual form of Bessel equation of non-integer order $\tilde{m}_\pm \equiv m \pm f$. It has the general positive-energy solution:

$$R_m(r) = F_m J_{|\tilde{m}_+|}(k_\perp r) + G_m J_{-|\tilde{m}_+|}(k_\perp r) \, ,$$

where $k_\perp$ is the component of $\vec{k}$ in the x-y plane.

Since $J_{\pm|n|}(x) \sim x^{\pm|n|}$ for $x \to 0$, $G_m = 0$ if we insist on the regularity at the origin. Physically speaking, the irregular solutions must be eliminated since the scalar particle carries no magnetic moment [13]. As a result, the particle suffers no interaction with the magnetic field at $\rho = 0$. In contrast, in the spinor case [13], there is an interaction between the spin and the magnetic moment.

Therefore, the wave functions in the latter case do not vanish at $r=0$. This leads to the problem of self-adjointness extension of the Hamilton operator [14-16].

The normalization constant $N$, can be determined by the following normalization condition [17]

$$\frac{i\hbar}{2mc^2}\int[\psi^*_{(\pm)}(\vec{r},t)\frac{\partial\psi_{(\pm)}(\vec{r},t)}{\partial t}-\psi_{(\pm)}(\vec{r},t)\frac{\partial\psi^*_{(\pm)}(\vec{r},t)}{\partial t}]d^3\vec{r}=1, \qquad (7)$$

within a cubic box of volume $V=L^3$.

However, the complete set of solutions of the interacting KG equation includes the negative-energy states, in addition to the positive-energy states. These states should be interpreted as antiparticles with positive-energy, which can be constructed from the negative-energy particle states by the charge-conjugation operation, in which $\psi_{(-)} \to \psi_c = \psi^*_{(-)}$ and replacing e by -e.

Consequently, the particle partial mode solution, for $E_n = E^a > 0,$ is given by

$$\psi_{(+)}(r,\varphi,z;t) = \sqrt{Mc^2/E_nV}\, e^{-i(c\,e_\hbar - k_3\,z - m\varphi)} J_{|\tilde{m}_+|}(k_\perp r), \qquad (8\text{-a})$$

while the antiparticle partial mode solution, for $E'_{\bar{n}} = E^{\bar{a}} > 0,$ is given by

$$\psi_c(r,\varphi,z;t) = \sqrt{Mc^2/E'_{\bar{n}}V}\, e^{i(c\,|e_{\bar{\pi}}|t - k_3\,z - m\varphi)} J_{|\tilde{m}_-|}(k_\perp r). \qquad (8\text{-b})$$

The cylindrical partial mode solutions do not describe *incoming* and *outgoing* particles with definite linear momenta at infinity. In order to find out the matrix-element, scattering solutions should be found based on the fact that the wave functions for the incoming and outgoing particles in the presence of the external AB field, should be written as

$$\Psi \stackrel{a}{=} \Psi_{out} = \sum_{m=-\infty}^{\infty} c_m \psi_m(\vec{r},t), \qquad (9\text{-a})$$

$$\Psi \stackrel{\bar{a}}{=} \Psi_{in} = \sum_{m'=-\infty}^{\infty} c_{m'} \psi_{m'}(\vec{r},t). \qquad (9\text{-b})$$

The coefficients $c_m$ are determined by using the fact that $\Psi_{in}$ must behave at large distance like a *plane wave* propagating in the direction $\vec{k}'$ plus an *outgoing cylindrical wave.* Likewise, $c_{m'}$ are determined since $\Psi_{out}$ must behave at large distance like a *plane wave* propagating in the direction $\vec{k}$ plus an *incoming cylindrical wave* .

By making use of the familiar expansion of plane waves in terms of Bessel functions together with the asymptotic form of Bessel functions, it can be shown that the amplitudes $c_m$ and $c_{m'}$ are given, respectively, by:

$$c_m = e^{i\{m(\varphi-\varphi_\perp) - \frac{\pi}{2}|\tilde{m}|\}} \quad \text{and} \quad c_{m'} = e^{i\{-m'(\varphi-\varphi'_\perp) + \frac{\pi}{2}|\tilde{m}'|\}} \qquad (10)$$

where $\varphi'_\perp$ and $\varphi_\perp$ are the angles of the direction of the incoming and outgoing momentum, respectively, as $r \to \infty$. With this choice of the coefficients for the incoming and outgoing particles, the interaction with the external AB field will be described in a correct way.

## III. The Matrix-Element Calculations

In order to calculate the matrix element for the production of a pair of massive particles by an incident high-energy massless particle such as the photon, it is necessary to find out first the probability amplitude given by [17]

$$\vec{d} = \frac{1}{2Mc} \int d^3 r\, e^{i\vec{k}\cdot\vec{r}} \left( \left[\hat{\vec{P}}^* \Psi^{a*}(\vec{r})\right] \Psi^{\bar{a}}(\vec{r}) + \Psi^{a*}(\vec{r}) \left[\hat{\vec{P}} \Psi^{\bar{a}}(\vec{r})\right] \right). \quad (11)$$

The integrand can be viewed as a coupling of the charged current of the scalar particles to the external radiation field. In this work, we followed the coordinate system defined in Fig. [1]. In this case,

$$\vec{k}\cdot\vec{r} = kr\sin(J_k)\cos(\varphi - \varphi_k) + k\cos(J_k)z.$$

$$\Psi^a(r,\varphi,z) = \sqrt{Mc^2/E_n V}\, e^{ik_3 z} \sum_{m=-\infty}^{m=\infty} c_m e^{im\varphi} J_{|m+f|}(k_\perp r)$$

$$\Psi^{\bar{a}}(r,\varphi,z) = \sqrt{Mc^2/E'_{\bar{n}} V}\, e^{-ik'_3 z} \sum_{m'=-\infty}^{m'=\infty} c_{m'} e^{-im'\varphi} J_{|m'-f|}(k'_\perp r).$$

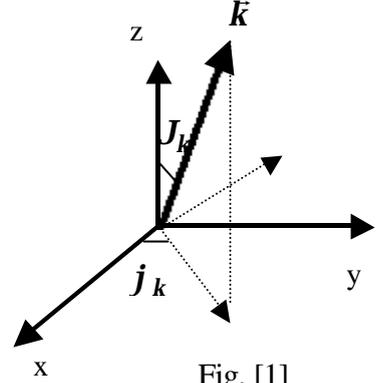

Fig. [1]

After extensive manipulations, it can be shown that

$$\vec{d} = d_1 e_1 + d_2 e_2 + d_z e_z \quad (12)$$

with

$$d_1 = \frac{-1}{4L^3 \sqrt{e_n e_{\bar{n}}}} \sum_{m=-\infty}^{\infty} \sum_{m'=-\infty}^{\infty} c_m c_{m'}^* \int_{-L/2}^{L/2} dz\, e^{i(k_3 - k_3 - k'_3)z} \int_0^\infty r\,dr \int_0^{2\pi} d\varphi\, e^{ik_\perp r \cos(\varphi - \varphi_k) - i(m+m')\varphi}$$

$$\{k'_\perp[(1+\text{sgn}(\tilde{m}'))[e^{-i(\varphi_k - \varphi)}Y(|\tilde{m}|,|\tilde{m}'|-1) + e^{i(\varphi_k - \varphi)}Y(|\tilde{m}|,|\tilde{m}'|+1)] - (1-\text{sgn}(\tilde{m}'))[e^{-i(\varphi_k - \varphi)}$$

$$Y(|\tilde{m}|,|\tilde{m}'|+1) + e^{i(\varphi_k - \varphi)}Y(|\tilde{m}|,|\tilde{m}'|-1)]] \;+\; k_\perp[(1-\text{sgn}(\tilde{m}))[e^{i(\varphi_k - \varphi)}Y(|\tilde{m}|-1,|\tilde{m}'|) +$$

$$e^{-i(\varphi_k - \varphi)}Y(|\tilde{m}|+1,|\tilde{m}'|)] - (1+\text{sgn}(\tilde{m}))[e^{i(\varphi_k - \varphi)}Y(|\tilde{m}|+1,|\tilde{m}'|) + e^{-i(\varphi_k - \varphi)}(|\tilde{m}|-1,|\tilde{m}'|)]\},$$

$$d_2 = \frac{i}{4L^3 \sqrt{e_n e_{\bar{n}}}} \sum_{m=-\infty}^{\infty} \sum_{m'=-\infty}^{\infty} c_m c_{m'}^* \int_{-L/2}^{L/2} dz\, e^{i(k_3 - k_3 - k'_3)z} \int_0^\infty r\,dr \int_0^{2\pi} d\varphi\, e^{ik_\perp r \cos(\varphi - \varphi_k) - i(m+m')\varphi}$$

$$\{k'_\perp[(1+\text{sgn}(\tilde{m}'))[-e^{-i(\varphi_k - \varphi)}Y(|\tilde{m}|,|\tilde{m}'|-1) + e^{i(\varphi_k - \varphi)}Y(|\tilde{m}|,|\tilde{m}'|+1)] - (1-\text{sgn}(\tilde{m}'))[e^{i(\varphi_k - \varphi)}$$

$$Y(|\tilde{m}|,|\tilde{m}'|-1) + e^{-i(\varphi_k - \varphi)}Y(|\tilde{m}|,|\tilde{m}'|+1)]] \;+\; k_\perp[(1-\text{sgn}(\tilde{m}))[e^{i(\varphi_k - \varphi)}Y(|\tilde{m}|-1,|\tilde{m}'|) -$$

$$e^{-i(\varphi_k - \varphi)}Y(|\tilde{m}|+1,|\tilde{m}'|)] + (1+\text{sgn}(\tilde{m}))[e^{-i(\varphi_k - \varphi)}Y(|\tilde{m}|-1,|\tilde{m}'|) - e^{i(\varphi_k - \varphi)}Y(|\tilde{m}|+1,|\tilde{m}'|)]\},$$

and

$$d_z = \frac{1}{2L^3 \sqrt{e_n e_{\bar{n}}}} \sum_{m=-\infty}^{\infty} \sum_{m'=-\infty}^{\infty} c_{m'} c_m^* \int_{-L/2}^{L/2} dz\, e^{i(k_3 - k_3 - k'_3)z} \int_0^\infty r\,dr \int_0^{2\pi} d\varphi\, e^{ik_\perp r \cos(\varphi - \varphi_k) - i(m+m')\varphi}$$

$$(k_3 - k'_3) Y(|\tilde{m}|, |\tilde{m}'|).$$

In the above expressions, the Y-functions are products of two Bessel functions:

$$Y(a,b) \equiv J_a(k_\perp r) J_b(k'_\perp r) \quad (13)$$

Calculation of the probability amplitude integral is a very involved task. First of all, making use of the box normalization formalism, in which integration over the z-coordinate is worked out from L/2 to L/2 instead of the usual space from $-\infty$ to $\infty$. Secondly, integration over $\varphi$-coordinate has the following general form:

$$\Xi = \int_0^{2\pi} d(\varphi - \varphi_\kappa) e^{-i\{(m+m'\pm 1)\varphi + \kappa\rho\sin\vartheta_\kappa \cos(\varphi-\varphi_\kappa)\}} .$$

It can be calculated by making use of the modified version of Sommerfeld representation of the Bessel function furnishing a third Bessel function. For more details, see Appendix A.

The result, with $\theta_\pm \equiv (m + m' \pm 1)$, is

$$\Xi = 2\boldsymbol{p} e^{-i\ q_\pm\ (j_k - \frac{p}{2})} J_{-q_\pm}(\boldsymbol{k}_\perp \boldsymbol{r}\ ) . \tag{14}$$

The remaining $\rho$-integrals are reduced to integrals over three Bessel functions of different orders and arguments that show algebraic relationships. They are of two types that can be solved by using tabulated formulae 6.578(3) and 6.522(14) of [18]:

$$\int_0^\infty x\ J_n(b_1 x) J_m(b_2 x) J_{m+n}(cx)\ dx = 0 \quad for: c > b_1 + b_2\ and\ b_1, b_2 > 0. \tag{15-a}$$

$$\int_0^\infty x\ J_m(cx\sin\boldsymbol{h}\cos\boldsymbol{z}) J_n(cx\cos\boldsymbol{h}\sin\boldsymbol{z}) J_{m-n}(cx) dx = \frac{2}{\boldsymbol{p}c^2}\sin(m\boldsymbol{p}) a^{\boldsymbol{m}}\ b^{\boldsymbol{n}} D, \tag{15-b}$$

$$for: c > 0,\ \text{Re}\ \boldsymbol{n} > -1,\ \boldsymbol{h} > 0,\ \boldsymbol{z} < \frac{\boldsymbol{p}}{2}$$

with

$$a \equiv \frac{\sin\boldsymbol{h}}{\cos\boldsymbol{z}},\ b \equiv \frac{\sin\boldsymbol{z}}{\cos\boldsymbol{h}},\ D \equiv [\cos(\boldsymbol{h}+\boldsymbol{z})\cos(\boldsymbol{h}-\boldsymbol{z})]^{-1}. \tag{16}$$

In this process, the total energy is conserved as well as the linear momentum along the magnetic tube (z-direction); only the radial momentum is not conserved and satisfies the relation

$$\boldsymbol{k}_\perp > k_\perp + k'_\perp, \tag{17}$$

which means that there is an excess of radial momentum, $\boldsymbol{k}_\perp - (k_\perp + k'_\perp)$, transmitted to the magnetic tube. Accordingly, we have to set

$b_1 = k_\perp,\ b_2 = k'_\perp,\ c = \boldsymbol{k}_\perp$ : For Eq.(15-a) type,

$c\cos\boldsymbol{h}\ \sin\boldsymbol{z} = k_\perp,\ c\sin\boldsymbol{h}\ \cos\boldsymbol{z} = k'_\perp,\ c = \boldsymbol{k}_\perp$ : For Eq.(15-b) type.

Thus, depending on the indices of Bessel functions, the integral can be directly solved, taking into account the conditions behind each integral type as well as the linear dependence properties of Bessel functions, for integer n, given by $J_{-n}(x) = (-1)^n J_n(x)$.

Then the sums in the probability amplitude components can be evaluated (after we redefine the indices such that $\bar{m} \equiv m + [f]$, $\bar{m}' \equiv m' - [f]$) since they reduce to geometric ones. An example of such a calculation is given in Appendix B.

Next, a deep insight into the expressions shows that the process turns out to be forbidden unless the redefined quantum numbers $\bar{m}$ and $\bar{m}'$ of the incoming and outgoing particles have opposite signs; mathematically speaking: $for\ \bar{m}' \geq 0,\ \bar{m} < 0\ and\ for\ \bar{m}' < 0,\ \bar{m} \geq 0$. Therefore, they have to satisfy the following **selection rule:**

$$\boxed{sign(\bar{m} * \bar{m}') = -1} \tag{18}$$

This means that the created charged particles need to pass the magnetic string in opposite direction. This is necessary for the ingoing photon to transmit the excess of its radial momentum to the string and create the real particle-antiparticle pair from the vacuum. This result was noticed before in the case of spin-1/2 particles [7], but it was never reported before in the case of scalar particles.

Next, summing over the redefined indices, the closed form expressions for the probability amplitude takes the following form:

$$\vec{d} = \frac{D\,d_{k_3,-k_3'} e^{i[f](\vec{j}_\perp' - \vec{j}_\perp)} \sin pd}{L^2 \sqrt{e_n e_{\bar{n}}}\, \mathbf{k}_\perp^2}[i\{A(e^{ipd}(ab)^d \Sigma^+ + e^{-ipd}(ab)^{-d}\Sigma^-)\}e_1 +$$

$$\{B(e^{ipd}(ab)^d \Sigma^+ - e^{-ipd}(ab)^{-d}\Sigma^-)\}e_2 + \{2(k_3 - k_3')(e^{ipd}(ab)^d \Sigma^+ - e^{-ipd}(ab)^{-d}\Sigma^-)\}e_z].$$
(19)

where

$$\Sigma^\pm \equiv \frac{1}{1-ae^{-i(\vec{j}_k - \vec{j}_\perp)}} \frac{be^{-i(\vec{j}_\perp' - \vec{j}_k)}}{1-be^{-i(\vec{j}_\perp' - \vec{j}_k)}}, \qquad \Sigma^{\overline{\equiv}} \frac{ae^{i(\vec{j}_k - \vec{j}_\perp)}}{1-ae^{i(\vec{j}_k - \vec{j}_\perp)}} \frac{1}{1-be^{i(\vec{j}_\perp' - \vec{j}_k)}}, \qquad (20)$$

$$A \equiv k_\perp(a - a^{-1}) + k_\perp'(b - b^{-1}), \qquad B \equiv k_\perp(a - a^{-1}) - k_\perp'(b - b^{-1}), \qquad (21)$$

$$a(k_\perp, k_\perp', \mathbf{k}_\perp) = \frac{2k_\perp \mathbf{k}_\perp}{(k_\perp'^2 + \mathbf{k}_\perp^2 - k_\perp^2) + \sqrt{k_\perp^4 - 2\mathbf{k}_\perp^2(k_\perp'^2 + k_\perp^2) + (k_\perp'^2 - k_\perp^2)^2}}, \qquad (22\text{-a})$$

$$b(k_\perp, k_\perp', \mathbf{k}_\perp) = \frac{2k_\perp' \mathbf{k}_\perp}{(k_\perp'^2 + \mathbf{k}_\perp^2 - k_\perp^2) + \sqrt{k_\perp^4 - 2\mathbf{k}_\perp^2(k_\perp'^2 + k_\perp^2) + (k_\perp'^2 - k_\perp^2)^2}}, \qquad (22\text{-b})$$

$$D = \frac{\mathbf{k}_\perp^2\, ab}{k_\perp k_\perp'[1 - a^2 b^2]}, \qquad (23)$$

$$\mathbf{j}_{ij} \equiv \mathbf{j}_i - \mathbf{j}_j. \qquad (24)$$

In Eq.(19), we considered only the case of normal incidence of the incoming photon on the solenoid, $\mathbf{J}_k = \mathbf{p}/2$. In this case, subsequent calculations become simpler without loss of generality.

For pair production process by a single photon of specific energy, the *transition probability per unit time* is given by [17]:

$$w_{\bar{n}n} = \frac{4p^2 e^2}{\hbar \mathbf{k} L^3} \cdot \sum_{n_3, n_3'} (\vec{d}^* \hat{a}^+)(\vec{d}\ \hat{a})\, d(e_{\bar{n}}\ e_n\ \mathbf{k}) \qquad (25)$$

where the sum is evaluated over the momentum z-component of the particles and antiparticles, respectively. Summation over $n_3$ and $n_3'$ can be performed with the help of the formula [17]

$$\sum_{n_3, n_3'} d_{k_3, -k_3'} = \sum_{n_3, n_3'} d_{n_3, -n_3'} = \frac{L}{2p} \int dk_3. \qquad (26)$$

It is clear that the created pair is emitted with momentum whose z-components are equal in absolute values but have opposite directions.

If the amplitudes are expanded in terms of the linear polarization vectors, $b_l$, described in Fig.[1], such that

$$\hat{b}_l = \begin{cases} \hat{e}^{(s)} = (-\sin\varphi_k, \cos\varphi_k, 0) & \text{for } l = 2 \\ \hat{e}^{(p)} = (-\cos\Theta_k \cos\varphi_k, -\cos\Theta_k \sin\varphi_k, \sin\Theta_k) & \text{for } l = 3 \end{cases} \quad (27)$$

then the transition probability per unit time for a given polarization state $\lambda$ reads

$$w_{\bar{n}n,\,l} = \frac{2\pi e^2}{\hbar k\, L^2} \int_{-\infty}^{\infty} dk_3'\ \Lambda_l\, \delta(e_{\bar{n}} + e_n - k) \quad (28)$$

with

$$\Lambda_{\bar{l}} \cdot (\vec{d}^*\ b_l)(\vec{d}\ b_l). \quad (29)$$

However, the calculations should be done with respect to the rotated coordinate system. Then, $e^{(s)} = (0, 1, 0)$ and $e^{(p)} = (-\cos\Theta_k, 0, \sin\Theta_k)$. In other words, the polarization vector may be taken to be *normal* to the plane determined by $\vec{k}$ and $e_2$ for $e^{(s)}$, while for $e^{(\pi)}$ it must then lie *in* that plane. But for normal incidence, it is not difficult to show that

$$\Lambda^{(s)} = \frac{D^2 \sin^2 pd}{L^4 e_n e_{\bar{n}} k_\perp^4} A^2 \{(ab)^{2d} |\Sigma^+|^2 + (ab)^{-2d} |\Sigma^-|^2 + e^{2ipd}\Sigma^+\Sigma^{-*} + e^{-2ipd}\Sigma^-\Sigma^{+*}\}. \quad (30\text{-a})$$

$$\Lambda^{(p)} = \frac{D^2 \sin^2 pd}{L^4 e_n e_{\bar{n}} k_\perp^4} 4(k_3 - k_3')^2 \{(ab)^{2d} |\Sigma^+|^2 + (ab)^{-2d} |\Sigma^-|^2 - e^{2ipd}\Sigma^+\Sigma^{-*} - e^{-2ipd}\Sigma^-\Sigma^{+*}\}, \quad (30\text{-b})$$

where

$$|\Sigma^+|^2 = \frac{b^2}{[1 - 2a\cos\varphi_{k\perp} + a^2][1 - 2b\cos\varphi'_{\perp k} + b^2]},\ |\Sigma^-|^2 = \frac{a^2}{[1 - 2a\cos\varphi_{k\perp} + a^2][1 - 2b\cos\varphi'_{\perp k} + b^2]}$$

Next, the complete information about energy, angular, and polarization distributions of created particles and antiparticles is contained in the effective differential cross section given by

$$d\sigma_l = \Gamma(e_n, e_{\bar{n}}) \frac{1}{J} dw_{n\bar{n}}. \quad (31)$$

where

$$\Gamma(e_n, e_{\bar{n}}) = \left(\frac{1}{c}\right)\left(\frac{L}{2\pi}\right)^4 k_\perp dk_\perp d\varphi_\perp k'_\perp dk'_\perp d\varphi'_\perp, \quad (32)$$

and

$$J = \frac{c}{L^2}. \quad (33)$$

Consequently,

$$\frac{d\sigma_l}{dk_\perp d\varphi_\perp dk'_\perp d\varphi'_\perp dk'_3} = \frac{2\pi e^2}{\hbar k}\left(\frac{L}{2\pi}\right)^4 k_\perp k'_\perp \Lambda_l\Big|_{e_n + e_{\bar{n}} = k}. \quad (34)$$

where $\Lambda_l$ are given in Eqs(30-a,b).

## IV. Limiting Cases:

The above differential cross-section is somewhat of complicated form. So it is worthwhile to illustrate limiting cases.

Non-relativistic Limit:

In this limit, the incident photon energy is just above the threshold energy. i.e. $\hbar w - 2Mc^2 \simeq 0$..
In this case, $\mathbf{k} \sim 2Mc/\hbar$ and $k_\perp \sim k'_\perp \sim k_3 \sim k'_3 \ll Mc/\hbar$. Then

$$a_{NR} \simeq \frac{k_\perp}{\mathbf{k}} = \frac{\hbar k_\perp}{2Mc}, \quad b_{NR} \simeq \frac{k'_\perp}{\mathbf{k}} = \frac{\hbar k'_\perp}{2Mc}, \quad D_{NR} \simeq 1, \quad A_{NR} \simeq -2\mathbf{k} = \frac{-4Mc}{\hbar}, \quad B_{NR} \simeq 0.$$

By substituting the above approximate values, the probability amplitude is considerably simplified. Its components read

$$d_1^{NR} \simeq \frac{-i\hbar^3 e^{i[f]\mathbf{j}_{\perp\perp}} \sin(\mathbf{pd})}{2L^2 M^3 c^3} \left\{ \left( \frac{\hbar^2 k_\perp k'_\perp}{4M^2 c^2} \right)^d k'_\perp e^{-i\mathbf{j}_{\perp k}+i\mathbf{pd}} + \left( \frac{\hbar^2 k_\perp k'_\perp}{4M^2 c^2} \right)^{-d} k_\perp e^{-i\mathbf{j}_{\perp k}-i\mathbf{pd}} \right\}, \quad (35\text{-a})$$

$$d_2^{NR} \simeq 0, \quad (35\text{-b})$$

$$d_z^{NR} \simeq \frac{\hbar^4 k_3 e^{i[f]\mathbf{j}_{\perp\perp}} \sin(\mathbf{pd})}{2L^2 M^4 c^4} \left\{ \left( \frac{\hbar^2 k_\perp k'_\perp}{4M^2 c^2} \right)^d k'_\perp e^{-i\mathbf{j}_{\perp k}+i\mathbf{pd}} + \left( \frac{\hbar^2 k_\perp k'_\perp}{4M^2 c^2} \right)^{-d} k_\perp e^{-i\mathbf{j}_{\perp k}-i\mathbf{pd}} \right\}. \quad (35\text{-c})$$

And

$$\Lambda_{NR}^{(s)} \simeq \frac{\hbar^6 \sin^2(\mathbf{pd})}{4L^4 M^6 c^6} \left\{ \left( \frac{\hbar^2 k_\perp k'_\perp}{4M^2 c^2} \right)^{2d} k'^2_\perp + \left( \frac{\hbar^2 k_\perp k'_\perp}{4M^2 c^2} \right)^{-2d} k^2_\perp + \left( \frac{\hbar^2 k_\perp k'_\perp}{4M^2 c^2} \right) \left[ e^{2\mathbf{pid}-i\mathbf{j}_{\perp\perp}} + e^{-2\mathbf{pid}+i\mathbf{j}_{\perp\perp}} \right] \right\},$$

$$\Lambda_{NR}^{(p)} \simeq \frac{\hbar^8 k_3^2 \sin^2(\mathbf{pd})}{4L^4 M^8 c^8} \left\{ \left( \frac{\hbar^2 k_\perp k'_\perp}{4M^2 c^2} \right)^{2d} k'^2_\perp + \left( \frac{\hbar^2 k_\perp k'_\perp}{4M^2 c^2} \right)^{-2d} k^2_\perp - \left( \frac{\hbar^2 k_\perp k'_\perp}{4M^2 c^2} \right) \left[ e^{2\mathbf{pid}-i\mathbf{j}_{\perp\perp}} + e^{-2\mathbf{pid}+i\mathbf{j}_{\perp\perp}} \right] \right\}.$$

Eventually,

$$\frac{d\mathbf{s}_{NR}^{(s)}}{dk_\perp d\mathbf{j}_\perp dk'_\perp d\mathbf{j}'_\perp dk'_3} = \frac{e^2 \hbar^5 k_\perp k'_\perp \sin^2(\mathbf{pd})}{(2\mathbf{p})^3 4\mathbf{k} M^6 c^6} \left\{ \left( \frac{\hbar^2 k_\perp k'_\perp}{4M^2 c^2} \right)^{2d} k'^2_\perp + \left( \frac{\hbar^2 k_\perp k'_\perp}{4M^2 c^2} \right)^{-2d} k^2_\perp + \right.$$

$$\left. \left( \frac{\hbar^2 k_\perp k'_\perp}{4M^2 c^2} \right) \left[ e^{2\mathbf{pid}-i\mathbf{j}_{\perp\perp}} + e^{-2\mathbf{pid}+i\mathbf{j}_{\perp\perp}} \right] \right\}, \quad (36\text{-a})$$

$$\frac{d\mathbf{s}_{NR}^{(p)}}{dk_\perp d\mathbf{j}_\perp dk'_\perp d\mathbf{j}'_\perp dk'_3} = \frac{e^2 \hbar^7 k_\perp k'_\perp k_3^2 \sin^2(\mathbf{pd})}{(2\mathbf{p})^3 4\mathbf{k} M^8 c^8} \left\{ \left( \frac{\hbar^2 k_\perp k'_\perp}{4M^2 c^2} \right)^{2d} k'^2_\perp + \left( \frac{\hbar^2 k_\perp k'_\perp}{4M^2 c^2} \right)^{-2d} k^2_\perp - \right.$$

$$\left. \left( \frac{\hbar^2 k_\perp k'_\perp}{4M^2 c^2} \right) \left[ e^{2\mathbf{pid}-i\mathbf{j}_{\perp\perp}} + e^{-2\mathbf{pid}+i\mathbf{j}_{\perp\perp}} \right] \right\}. \quad (36\text{-b})$$

A remarkable feature can be deduced that in the nonrelativistic limit, the differential scattering cross section for the $\mathbf{s}$-polarization is much larger than the $\mathbf{p}$-polarization. This means that it is unlikely that the $\mathbf{s}$-polarized photon creates a scalar particle-antiparticle pair.

B. Ultrarelativistic Limit:

If the photon energy is much larger than the threshold energy, i.e. $c\hbar k \gg 2Mc^2$, then the created pair will be emitted predominantly in the forward direction within too narrow cone, about the direction of the incident photon. This is because the pair energy is approximately of kinetic type, in the lowest order approximation, i.e. $e_n \simeq k \equiv \sqrt{k_3^2 + k_\perp^2}$, $e_{\bar{n}} \simeq k' \equiv \sqrt{k'^2_3 + k'^2_\perp}$.

Accordingly, it is not difficult to conclude that $\varphi_\kappa \cong \varphi_\perp \cong \varphi_{\perp'}$. Then the angular distribution of the emitted pair is simplified considerably. In this limit

$$\Sigma_{UR}^+ = \Sigma_{UR}^{+*} \simeq \frac{b_{UR}}{(1-b_{UR})(1-a_{UR})}, \qquad (37\text{-a})$$

$$\Sigma_{UR}^- = \Sigma_{UR}^{-*} = \frac{a_{UR}}{b_{UR}} \Sigma_{UR}^+ \simeq \frac{a_{UR}}{(1-b_{UR})(1-a_{UR})}. \qquad (37\text{-b})$$

Therefore,

$$\Lambda_{UR}^{(s)} \simeq \frac{D_{UR}^2 \sin^2 \boldsymbol{p}\boldsymbol{d}}{L^4 k k' \boldsymbol{k}_\perp^4} A_{UR}^2 \left(\Sigma^+\right)^2 \{(a_{UR}b_{UR})^{2d} + a_{UR}^2 b_{UR}^{-2}(a_{UR}b_{UR})^{-2d} + 2a_{UR}b_{UR}^{-1}\cos 2\boldsymbol{p}\boldsymbol{d}\},$$

$$\Lambda_{UR}^{(p)} \simeq \frac{4D_{UR}^2 (k_\perp - k'_\perp)^2 \sin^2 \boldsymbol{p}\boldsymbol{d}}{L^4 k k' \boldsymbol{k}_\perp^4} \left(\Sigma^+\right)^2 \{(a_{UR}b_{UR})^{2d} + a_{UR}^2 b_{UR}^{-2}(a_{UR}b_{UR})^{-2d} - 2a_{UR}b_{UR}^{-1}\cos 2\boldsymbol{p}\boldsymbol{d}\}.$$

where $a_{UR}$ and $b_{UR}$ are the ultrarelativistic versions of Eqs.(22-a,b).

*V. Conclusions*

We have analysed the scalar pair production by a single, high-energy linearly polarized photon in the presence of AB potential. In this case, photons do not intertact directly with the magnetic field because they cannot penetrate the magnetic string, and the process hasppens due to the interaction of the created charged particles with the AB vector potential.

The differetial scattering cross section, was evaluated exactly, in the framework of time-tependent perturbation theory. We showed that the process turns out to be forbidden unless the quantum numbers $\bar{m}$ and $\bar{m}'$ of the the outgoing particle and antiparticle, respectivley, have opposite signs. This means that the created virtual pair to encircle the magnetic string in opposite directions, in order to be transformed into a real one. It is interesting to point out that the process seems to be somewhat mystrious because it results dut to a global and topological reason only. So the mechanism responsible for its occurrence resembles that process which takes near a cosmic string.

# Appendix A

The $\varphi$-integral has the following general form:

$$\Xi = \int_0^{2\pi} d(\varphi - \varphi_k)\ e^{-i\{(m+m'\pm)(\varphi - \varphi_k) - k_\perp r \cos(\varphi - \varphi_k)\}}. \quad \text{(A-1)}$$

Let $\varphi - \varphi_k = c + \pi$, then with $q_\pm \equiv m + m' \pm 1$ and $\bar{\varphi}_k \equiv \varphi_k - \dfrac{\pi}{2}$,

$$\Xi = e^{-i q_\pm (\varphi_k - \pi)} \int_{-\pi}^{\pi} dc\ e^{i q_\pm c + i k_\perp r \cos c}. \quad \text{(A-2)}$$

of order $\nu$ given by:

$$J_n(z) = \frac{e^{i\pi n/2}}{2\pi} \int_C dt\, e^{-iz\cos t + int} \quad \text{(A-3)}$$

where the contour C goes from $-\pi + \chi + i\infty$ to $\pi + \chi + i\infty$ with $\chi$ a positive infinitesimal as illustrated in Fig. [2].

Thus the final result of the $\varphi$-integral takes the following form:

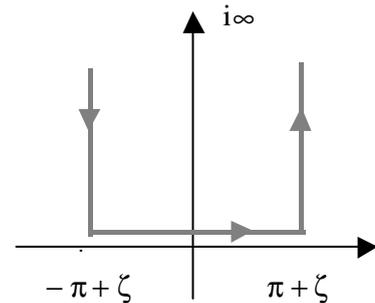

Fig. [2]

$$\Xi = 2p\, e^{-i\, q_{\pm}\, (j_k - \frac{p}{2})} J_{-q_{\pm}}(k_{\perp} r)\ . \tag{A-4}$$

# Appendix B

In this Appendix, we show a sample calculation of the first $\rho$-integral in $d_1$ expression, as an example.

Let

$$G_1 = \sum_{m=-\infty}^{\infty} \sum_{m'=-\infty}^{\infty} c_{m'} c_m^* e^{-i(m+m')\bar{J}_k} \left[\mathrm{sgn}(\tilde{m})-1\right] \int_0^{\infty} r\, dr\, J_{|\tilde{m}|-1}(k_{\perp} r) J_{|\tilde{m}'|}(k'_{\perp} r) J_{q_+}(k_{\perp} r) \tag{B.1}$$

First of all, the sum indices m and m′ should be redefined as follows:

$$\begin{cases} \bar{m} = m + [f] \Rightarrow \quad = \tilde{m} + \bar{m}\ d \\ \bar{m}' = m' - [f] \Rightarrow \quad \tilde{m}' \quad \bar{m}'\ d \end{cases} \Rightarrow m + m' = \bar{m} + \bar{m}'. \tag{B.2}$$

Next, summing over the new indices and partitioning both summations for positive and negative indices values, then $G_1$ should be partitioned into four parts or terms. We will designate these terms with the letters $T_1$, $T_2$ $T_3$ and $T_4$, respectively.

$$G_1 = \sum_{\bar{m}=-\infty}^{-1} \sum_{\bar{m}'=-\infty}^{-1} \mathsf{A}(\bar{m},\bar{m}') + \sum_{\bar{m}=-\infty}^{-1} \sum_{\bar{m}'=0}^{\infty} \mathsf{A}(\bar{m},\bar{m}') + \sum_{\bar{m}=0}^{\infty} \sum_{\bar{m}'=-\infty}^{-1} \mathsf{A}(\bar{m},\bar{m}') + \sum_{\bar{m}=0}^{\infty} \sum_{\bar{m}'=0}^{\infty} \mathsf{A}(\bar{m},\bar{m}') \tag{B3}$$

with

$$\mathsf{A}(\bar{m},\bar{m}') \equiv c_{\bar{m}}^* c_{\bar{m}'} e^{-i(\bar{m}+\bar{m}')\bar{J}_k} \left[\mathrm{sgn}(\bar{m}+d)-1\right] \int_0^{\infty} r\, dr\, J_{|\tilde{m}|-1}(k_{\perp} r) J_{|\tilde{m}'|}(k'_{\perp} r) J_{q_+}(k_{\perp} r). \tag{B.4}$$

<u>First Term: $T_1$</u>

$$T_1 = \sum_{\bar{m}=-\infty}^{-1} \sum_{\bar{m}'=-\infty}^{-1} c_{\bar{m}}^* c_{\bar{m}'} e^{-i(\bar{m}+\bar{m}')\bar{J}_k} \left[\mathrm{sgn}(\bar{m}+d)-1\right] \int_0^{\infty} r\, dr\, J_{|\bar{m}+d|-1}(k_{\perp} r) J_{|\bar{m}'-d|}(k'_{\perp} r)\ J_{\bar{m}+\bar{m}'+1}(k_{\perp} r).$$

For $\bar{m} < 0$, $\bar{m}' < 0$: $|m+d| = -\bar{m}-d$, $|\bar{m}'-d| = -\bar{m}'+d$.

Making use of the fact that $J_{-n}(x)$ and $J_n$

$$T_1 = 2 e^{i[f](j_{\perp}' - j_{\perp})} \sum_{\bar{m}=-\infty}^{-1} e^{i\bar{m}(j_{\perp} - j_k - p/2)} (-1)^{-\bar{m}} \sum_{\bar{m}'=-\infty}^{-1} e^{i\bar{m}'(j_{\perp}' - j_k - p/2)} (-1)^{-\bar{m}'}$$

$$\int_0^{\infty} dr\, r\, J_{-\bar{m}-d-1}(k_{\perp} r) J_{-\bar{m}'+d}(k'_{\perp} r) J_{-\bar{m}-\bar{m}'-1}(k_{\perp} r).$$

A deep insight on the integral in $T_1$ expression shows that it must vanish since it is an integral of the Eq.(15-a) type. This is because
$(-\bar{m} - d - 1) + (-\bar{m}' + d) = (-\bar{m} - \bar{m}' - 1)$ for $k_{\perp} > k_{\perp} + k'_{\perp}$.

Second Term: $T_2$

$$T_2 = \sum_{\overline{m}=-\infty}^{-d} \sum_{\overline{m}'=0}^{\infty} c_{\overline{m}}^* c_{\overline{m}'} e^{-i(\overline{m}+\overline{m}')\bar{J}_k} \left[\text{sgn}(\overline{m}+d)-1\right] \int_0^{\infty} r dr \, J_{|\overline{m}+d|-1}(k_\perp r) J_{|\overline{m}'-d|}(k'_\perp r) \, J_{\overline{m}+\overline{m}'+1}(\mathbf{k}_\perp \mathbf{r}).$$

It is not difficult to prove that the above integral, for $\overline{m}<0$ & $\overline{m}'>0$, is of type Eq.(15-b) with $\mathbf{n}=-\overline{m}-\mathbf{d}-1$ and $\mathbf{m}=\overline{m}'-\mathbf{d}$ for $\mathbf{k}_\perp > k_\perp + k'_\perp$. Thus

$$T_2 = \frac{-4D}{p \, k_\perp^2} a^{-d-1} b^{-d} e^{i[f](\mathbf{j}'_\perp-\mathbf{j}_\perp)} e^{-ipd} \sum_{\overline{m}=-\infty}^{-1} a^{-\overline{m}} e^{i\overline{m}(\mathbf{j}_\perp-\mathbf{j}_k-p)} \sin[p(m+d+1)] \times$$

$$\sum_{\overline{m}'=0}^{\infty} b^{\overline{m}'} e^{i\overline{m}'(\mathbf{j}'_\perp-\mathbf{j}_k)}$$

Notice that the two summations are turned out to be geometric.

$$\sum_{\overline{m}'=0}^{\infty} b^{\overline{m}'} e^{i\overline{m}'(\mathbf{j}'_\perp-\mathbf{j}_k)} = \frac{1}{1-be^{i(\mathbf{j}'_\perp-\mathbf{j}_k)}}$$

Likewise,

$$\sum_{\overline{m}=-\infty}^{-1} a^{-\overline{m}} e^{i\overline{m}(\mathbf{j}_\perp-\mathbf{j}_k-p)} \sin[p(m+d+1)] = \sum_{\overline{m}=-\infty}^{-1} a^{-\overline{m}} e^{i\overline{m}(\mathbf{j}_\perp-\mathbf{j}_k)} \left(\frac{e^{ipd}-e^{-ipd}}{2i}\right)$$

$$= \sum_{\overline{\ell}=1}^{\infty} a^{\overline{\ell}} e^{-i\overline{\ell}(\mathbf{j}_\perp-\mathbf{j}_k)} \sin pd = \frac{\sin pd \, ae^{-i(\mathbf{j}_\perp-\mathbf{j}_k)}}{1-ae^{-i(\mathbf{j}_\perp-\mathbf{j}_k)}}$$

where we have set $\overline{\ell} \equiv -\overline{m}$. Finally

$$T_2 = G_1 = \frac{-4D \sin pd}{p k_\perp^2} e^{i[f](\mathbf{j}'_\perp-\mathbf{j}_\perp)} (ab)^{-d} e^{-ipd} \frac{e^{-i(\mathbf{j}_\perp-\mathbf{j}_k)}}{1-ae^{-i(\mathbf{j}_\perp-\mathbf{j}_k)}} \frac{1}{1-be^{i(\mathbf{j}'_\perp-\mathbf{j}_k)}}. \tag{B.5}$$

Third Term: $T_3$

$$T_3 = \sum_{\overline{m}=0}^{\infty} \sum_{\overline{m}'=-\infty}^{-1} c_{\overline{m}}^* c_{\overline{m}'} e^{-i(\overline{m}+\overline{m}')\overline{\varphi}_\kappa} \left[\text{sgn}(\overline{m}+\delta)-1\right] \int_0^{\infty} \rho d\rho \, J_{|\overline{m}+\delta|-1}(k_\perp \rho) J_{|\overline{m}'-\delta|}(k'_\perp \rho) J_{\overline{m}+\overline{m}'+1}(\kappa_\perp \rho).$$

For $\overline{m}>0$, $\overline{m}'<0: |\overline{m}+\mathbf{d}|=\overline{m}+\mathbf{d}$, $|\overline{m}'-\mathbf{d}|=-\overline{m}'+\mathbf{d}$. Also $\text{sgn}(\overline{m}+\mathbf{d})=1$. Therefore, the third term $T_3$ vanishes.

Fourth Term: $T_4$

$$T_4 = \sum_{\overline{m}=0}^{\infty} \sum_{\overline{m}'=0}^{\infty} c_{\overline{m}}^* c_{\overline{m}'} e^{-i(\overline{m}+\overline{m}')\bar{J}_k} \left[\text{sgn}(\overline{m}+d)-1\right] \int_0^{\infty} r dr \, J_{|\overline{m}+d|-1}(k_\perp r) J_{|\overline{m}'-d|}(k'_\perp r) J_{\overline{m}+\overline{m}'+1}(\mathbf{k}_\perp \mathbf{r}).$$

For $\overline{m}>0$, $\overline{m}'>0: |\overline{m}+\mathbf{d}|-1=\overline{m}+\mathbf{d}-1$, $|\overline{m}'-\mathbf{d}|=\overline{m}'-\mathbf{d}$. Also $\left[\text{sgn}(\overline{m}+\mathbf{d})-1\right]=0$.
Therefore, the fourth term $T_4$ vanishes too.